Title: INFORMATION SYSTEMS DEVELOPMENT AND EVOLUTION: A REPLICATION STUDY ON WORK DISTRIBUTION IN NORWEGIAN ORGANIZATIONS

Corresponding Author: Prof. John Krogstie, PhD

Corresponding Author's Institution: NTNU

First Author: Tor Kristian Veld

Order of Authors: Tor Kristian Veld; John Krogstie, PhD

Abstract: The information systems landscape is at first sight very different from how it was 20 years ago. On the other hand, it seems that we are still struggling with many of the same problems, including late or abandoned projects and unfilled customer demands. In this article we present selected data from survey investigations from 1993, 1998, 2003, 2008 and 2013 among Norwegian organizations on how they conduct information systems development and maintenance. In particular we compare data from 2008 and 2013in more detail. A major finding is that whereas main work distribution numbers was very stable between 2003 and 2008, we see some changes as for time used on maintenance and development between 2008 and 2013. Even if we witness large changes in the underlying implementation technology and methods used, a number of aspects such as application portfolio upkeep (the amount of work for keeping the application portfolio operational) though are still on the same level as it has been the last 15 years. On the other hand, because of the more complex infrastructures supporting the application portfolio, and the increasing number of in particular external users, an increasing amount of resources is used for other tasks such as operations and user-support than in the first investigations, although also this appears to have stabilised between the last investigations.

# INFORMATION SYSTEMS DEVELOPMENT AND EVOLUTION: A REPLICATION STUDY ON WORK DISTRIBUTION IN NORWEGIAN ORGANIZATIONS

Abstract

*The information systems landscape is at first sight very different from how it was 20 years ago. On the other hand, it seems that we are still struggling with many of the same problems, including late or abandoned projects and unfilled customer demands. In this article we present selected data from survey investigations from 1993, 1998, 2003, 2008 and 2013 among Norwegian organizations on how they conduct information systems development and maintenance. In particular we compare data from 2008 and 2013in more detail. A major finding is that whereas main work distribution numbers was very stable between 2003 and 2008, we see some changes as for time used on maintenance and development between 2008 and 2013. Even if we witness large changes in the underlying implementation technology and methods used, a number of aspects such as application portfolio upkeep (the amount of work for keeping the application portfolio operational) though are still on the same level as it has been the last 15 years. On the other hand, because of the more complex infrastructures supporting the application portfolio, and the increasing number of in particular external users, an increasing amount of resources is used for other tasks such as operations and user-support than in the first investigations, although also this appears to have stabilised between the last investigations.*

## 1 INTRODUCTION

Large changes in how we develop information systems and the underlying technology for information systems have been witnessed over the last decades.  For instance, over this period the prevalent development methods, programming languages and underlying technological infrastructure have changed substantially. In the early nineties, one moved from mainframe solutions to a client-server, and then to an internet and multi-channel architecture for many applications. Year 2000 and the dot.com situation had large impact at least temporarily on the development and maintenance of systems. More lately Service-oriented Architecture (SOA), outsourcing, cloud computing, mobile technologies and agile development methods have become popular and would be expected to have impact on information systems support effectiveness in organizations. According to (Jones 2006) one of the impacts on IS-development is the increasing amount of time used for maintenance of systems (instead of developing new systems).  On the other hand, many of the intrinsic problems and aspects related to information systems support in organizations are similar now to what they were 20 years ago. Application systems are valuable when they provide information in a manner that enables people to meet their evolving objectives more effectively (Boehm & Sullivan 1999). Many have claimed that the large amount of work that goes into maintenance compared to the amount of work used for development is a sign on poor use of resources to meet these demands. On the other hand, as stated already in (Brooks 1987), it is one of the essential properties of application systems that they are under a constant pressure of change, and thus should change to stay relevant. Given the evolutionary nature of the sources of system demands, it should come as no surprise that specifications and the related information system must evolve as well (Boehm & Sullivan 1999).

The goal of both development activities and maintenance activities is to keep the overall information system support of the organization relevant to the organization, meaning that the systems support the fulfilment of organizational goals.  A lot of the activities labelled 'maintenance', so-called enhancive maintenance, are in this light value-adding activates, enabling the users of the systems to do new task. On the other hand, a large proportion of the 'new' systems being developed are replacement systems, primarily replacing the existing systems without adding much to what end-users can do with the overall application systems portfolio of the organization.

Based on this argumentation we have earlier developed the concept application portfolio evolution (Krogstie 1995) as a more meaningful high-level measure to evaluate to what extent an organization is able to evolve their application system portfolio efficiently. How application portfolio evolution is different from traditional development is described further in the next section.

In this paper, we present descriptive results from two survey-investigations performed in Norwegian organizations during the end of 2013/early 2014, comparing with similar investigations done in Norway in 2008, 2003, 1998 and 1993. These investigations are also comparable to similar investigation by Lientz and Swanson going back to the late 70ties, thus are able to act as replication studies (Brooks et al. 2008) giving us a way of tracking the developments over the last 30 years in this area.

We will first give definitions of some of the main terms used within software development and maintenance, including the terms application portfolio upkeep and application portfolio evolution. We describe the research method including potential threat to validity, before the main descriptive results from our investigations are presented and compared with previous investigations from earlier years. Section 5 investigates in more detail stated hypothesis on the development trends on distribution of work on development and maintenance activities. The last section summarizes our results and presents possible future work.

## 2 BASIC CONCEPTS

Maintenance has traditionally been divided into three types: corrective, adaptive and perfective (IEEE 1999) inspired by e.g. (Swanson 1976). This vocabulary is well established both in theory and practice, and we here use the IEEE terms with some clarifications and further division also anchored in the literature:

Maintenance is defined as the process of modifying a software system or component after initial delivery to production.

1. Corrective maintenance is work done to correct faults in hardware and software.
2. Adaptive maintenance is work done to make the computer program usable in a changed environment.
3. Perfective maintenance is work done to improve the performance, maintainability, or other attributes of a computer program. Perfective maintenance has been divided into enhancive maintenance (Chapin 2000) and non-functional perfective maintenance. *Enhancive* maintenance involves changes and additions to the functionality offered to the users by the system. Non-functional perfective maintenance implies improvements to the quality features of the information system and other features being important for the developer and maintainer of the system, such as modifiability. Non-functional perfective maintenance thus includes what is termed preventive maintenance, but also such things as improving the performance of the system.

In addition to the traditional temporal distinction between development and maintenance, we have introduced the concepts application portfolio evolution and application portfolio upkeep (originally termed functional development and functional maintenance when originally introduced in (Krogstie 1995)).

1. Application portfolio upkeep: Work made to keep the functional coverage of the information system portfolio of the organization at the current level. This includes:

    a) Corrective maintenance
    b) Adaptive maintenance
    c) Non-functional perfective maintenance
    d) Development of replacement systems

2. Application portfolio evolution: Development or maintenance where changes in the application increase the functional coverage of the total application systems portfolio of the organization. This includes:

   a) Enhancive maintenance
   b) Development of new systems that cover areas, which are not covered earlier by other systems in the organizations

*Figure 1. Different types of development and maintenance activities*

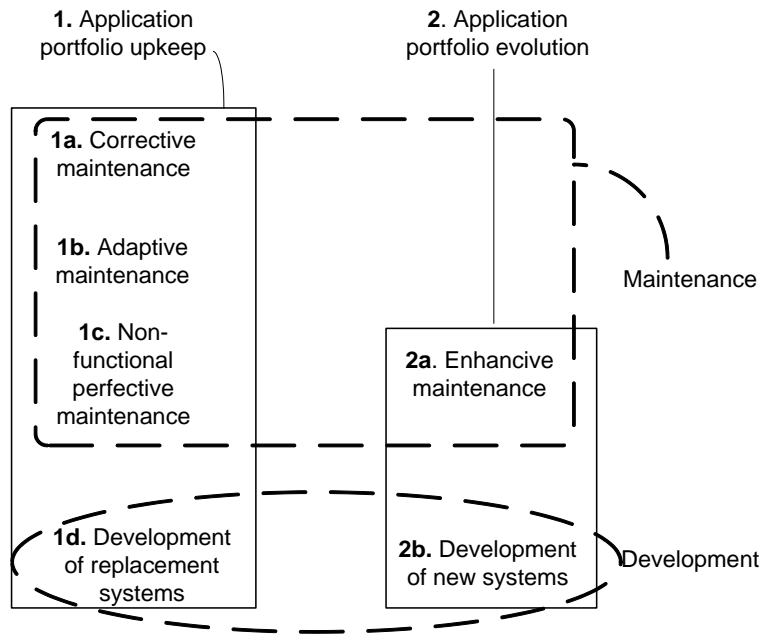

The relationships between the different concepts, using the same numbering as above are illustrated in Figure 1.

We note that some writers provide more detailed overview of maintenance tasks (Chapin et al 2001, Jones 2006). Jones (2006) has in total 21 categories: This includes user-support as a part of maintenance (a view shared with e.g. Dekleva (1992)), an area usually looked upon as belonging to 'other work' in other investigations, including ours.

## 3 RESEARCH METHOD

In connection to this work, we have performed two surveys. One is our main replication study covering a large number of topics matching the ones we have investigated 4 times earlier. The other is in connection to the yearly 'IT i praksis' (Rambøll 2014) investigation done early 2014, where we have included the questions relative to work distribution from our replication study, to compare this with e.g. the benefit of IT, alignment of IT and business strategy. We present these investigations individually below.

Our main replication survey was implemented in the SurveyMonkey web-tool and invitations were distributed by e-mail to 388 Norwegian organizations. The organizations were randomly selected from the list of member organizations of DnD (The Norwegian Computer Society - NCS) (NCS has currently around 1000 member organizations primarily in the private sector) and OSDF - the public sector IT-forum, to have also respondents from the public sector. IT i praksis was sent out to more than 500 organizations, equally divided between public and private sector, and we made sure to avoid overlap between the respondents.

The form in our replication study contained 41 questions including demographic data. The contents of the form were based on previous investigations within this area; especially those described in (Davidsen & Krogstie 2010, Holgeid, Krogstie & Sjøberg 2000, Krogstie & Sølvberg 1994, Krogstie, Jahr & Sjøberg 2003, Lientz & Swanson 1980, Nosek & Palvia 1990, Swanson & Beath 1989). The full survey form is available on request (main parts of this are available in the Appendix). Data from the replication study can also be made available. In this paper, we present both data from the replication study alone, and result joining the data from the replication study with the IT i Praksis study.

On some of the questions, we were interested in the quality of the answers, recognizing that some of the information called for might not be easily obtainable for all. It was also room for issuing open-ended remarks.

Galtung (1967) regards that the least size that is meaningful in a survey is 40 units. Since survey-investigations in the area of development and maintenance of application systems toward the same population earlier had given a response rate in the area of 22%-28% (Davidsen & Krogstie 2010) and the response rate of similar surveys has been around 20-25% (e.g. (Lientz & Swanson 1980, Nosek & Palvia 1990)), an answer ratio of approximately 20% was expected. This would have resulted in around 77 responses. 87 responses were returned, giving a response rate of 22%. Out of these only 68 responses could be used for the analysis. The additional responses were not complete, and in particular did not include responses to the questions relative to distribution of work.

Although this provides sufficiently many responses to provide grounds for doing statistical analysis, it would obviously be better to have a larger number of responses, based on a higher response rate. Our other investigation, linked to 'IT i praksis' supported this goal. Out of 533 distributed survey forms in this investigation, 272 responded (i.e. 51%) where returned, although only 208 provided responses to the questions matching our main replication study. When we put the results together, we had a total response-rate of 39%.

The forms in our main investigation were filled in using the web-form by people with long experience with application systems related work (average 21, 3 years), typically filling the role as IT director in the company. Judged on the responses, all organizations where doing work on all support-line levels (1-3) (Kajko-Mattson 2004), but with different emphasis on different types of support, and different patterns of (out) sourcing of activities. Because of this we will be cautious in our interpretations of the results.

### 3.1. Previous investigations

Being a replication study, we will compare some of the results with the results of similar investigations done before. The most important of these investigations are:

1. The investigation carried out by Davidsen and Krogstie in 2008 reported in (Davidsen & Krogstie, 2010a)

2. The investigation carried out by Jahr, Krogstie, and Sjøberg in 2003 reported in (Krogstie, Jahr & Sjøberg 2006).

3. The investigation carried out by Holgeid, Krogstie and Sjøberg in 1998 reported in (Holgeid, Krogstie & Sjøberg 2000).

4. The investigation carried out by Krogstie in 1993 reported in (Krogstie & Sølvberg 1994).

5. The Lientz and Swanson investigation (LS) reported in (Lientz & Swanson 1980): That investigation was carried out in 1977, with responses from 487 American organizations on 487 application systems (one system per company).

6. The Nosek and Palvia investigation (NP) reported in (Nosek & Palvia 1990): A follow-up study to Lientz/Swanson performed in 1990 asking many of the same questions as those of LS. Their results are based on responses from 52 American organizations.

7. The Swanson and Beath investigation (SB) (Swanson & Beath 1989): Reports on in-depth case-studies of 12 American companies that in addition to questions given in the Lientz/Swanson study focused on portfolio analysis and the development of replacement systems. These aspects are also a major part of our investigation.

The four first surveys in the list are the main Norwegian investigations in the Lientz/Swanson tradition. They contain the results from investigations of 77, 54, 52 and 53 Norwegian organizations, respectively. Special papers have in addition investigated specific aspects, including use of methodology and tools (Krogstie 1996), differences between private and public sector (Krogstie 2012), and comparisons of long-term trends (Krogstie 2006, Davidsen & Krogstie 2010b). In addition to these investigations, a number of later investigations have been done, but they typically focus on the distribution of maintenance tasks only (Gupta et al. 2006, Lee & Jefferson 2005, Mohaghegdi & Conradi 2004, Schach et al 2003), many only looking on the situation in one organization or on one application.

Several of the organizations that received a survey-form in the 1993, 1998, 2003 and 2008 studies also received the invitation to fill out the form in 2013, and many of the same questions have been asked. The methods that are used are also similar, enabling us to present a replication study, although the overlap among actual respondents to the survey is limited to only a few organizations across different instalments of the survey. Even if the population selection process has been similar, the actual organizations in these populations have changed a lot over the period of twenty years, both because of changed focus on IT, and because of the volatile business environment, with a number of acquisitions, mergers and bankruptcies.

## 3.2 Threats to Validity

The results of our study should be interpreted cautiously as there are several potential threats to its validity. The discussion below is based on recommendations given in (Jørgensen 1994, Kitchenham et al 2002).

### 3.2.1 Population

The sample of our study was initially intended to represent the population of Norwegian companies or organizations with own development and maintenance work. Since a substantial number of the major Norwegian IT companies are members of The Norwegian Computer Society (NCS) we pragmatically chose the around 1000 member companies of NCS as our population for companies in the private sector. To have respondents also from the public sector, we chose the member organizations of OSDF - the public sector IT-forum. We distributed our survey forms to a random selection of 388 of those companies. 'IT i praksis' which is also done in collaboration with NCS is sent to both private and public sector organizations that perform at least part of the IT-activities internally. There are currently more than 800 organizations in public sector, only some of which are members of OSDF. Some of the large public sector organizations are also members of NCS, thus the total population is roughly close to 2000 organizations. Other studies also use member lists as a source of subjects, e.g. (Lientz & Swanson 1980). In particular, NCS-members were also used in the studies in 2008, 2003, 1998 and 1993. As noted above, the actual responding organizations have changed a lot between the different studies which is inevitable in such studies, thus rather than presenting a truly longitudinal study, what we are able to do is to present a replication study.

### 3.2.2 Response rate

According to (Saunders, Lewis & Thornhill 2009), it is common for Internet and e-mail surveys with a response rate of 11 % or lower. The response rate on the main replication survey of 22% can still be argued to be rather low. We experienced the same kind of problem of getting a high response rate in 2008, 2003, 1998 and 1993. The 'IT i prak-

sis'-investigation though had a response rate of 51%, and as reported we had around 39% responded in total. These close to 300 organizations are only around 15% of the total population described above though. A problem with a low response rate is that the respondents may not be representative of the population, that is, the companies may be particularly mature, have less pressure (they have time to answer survey forms), etc. However, the same selection mechanism was used in the 2013, 2008, 2003, 1998 and 1993 studies, so a comparison between those five studies to identify trends should be reasonable.

### 3.2.3 Respondents

Most of the persons who responded were IT managers in the company. They may have different views of the reality than developers and maintainers. For example, Jørgensen (1994) found that manager estimates of the proportion of effort spent on corrective maintenance were biased towards too high values when based on best guesses instead of good data, see also (Schach et al 2003) which report a similar effect. There might be biases in our study of this kind, but they may not affect the comparison with the 2008, 2003, 1998 and 1993 studies as the survey forms of these studies were also filled in by IT managers. Also in the previous American studies, IT managers have responded to the surveys.

### 3.2.4 Understanding of concepts

Achieving consistent answers requires that the respondents have a common understanding of the basic concepts of the survey form. This may be difficult to ensure in practice. For example, Jørgensen (1994) found that the respondents used their own definition of, for example, "software maintenance" even though the term was defined at the beginning of the questionnaire. We conducted a pilot study in a few companies to detect unclear questions and whether the time for filling-in the forms was reasonable. On earlier versions of the form we have done similar pilots and also got comments from several colleagues including experts in cognitive psychology which were highly familiar with the use of survey techniques and ensuring clarity of the formulation of questions. The forms were then refined. For many questions, there was space available to issue comments. This possibility together with the possibility to crosscheck numbers between different questions, were the main mechanisms used to identify possible misunderstanding among the respondents, which could be followed up afterwards.

### 3.2.5 Biased questions

Among the risks when designing survey forms are leading or sensitive questions, resulting in biased or dishonest answers. We believe that we have mostly avoided this problem. We promised and effectuated full confidentiality to the respondents.

### 3.2.6 Quality of data

On some of the questions, we were interested in the quality of the answers, recognizing that some of the information called for might not be easily obtainable. Answers of some of the quantitative questions were checked against each other for control. The remarks made on the questions gave more insight into the answers. We qualified for instance all data regarding distribution of work both in our study and the studies in 1993, 1998, 2003 and 2008 without finding significant differences on the variables we have used in the hypothesis testing between those reporting having good data and those reporting qualified guesses.

# 4 DESCRIPTIVE RESULTS

First, we present some of the overall demographics of the survey. Similar results from our previous surveys conducted in 2008, 2003, 1998, and 1993 are included in parenthesis where the numbers are comparable.

50% (2008- 40%, 2003-20%; 1998-43%) of the organizations had a yearly data processing budget above 10 mill NKr (approx. 1.3 mill USD), and the average number of employees among the responding organizations was 1083 (2008-1083; 2003-181; 1998-656; 1993-2347). The average number of full-time personnel in the IS-organizations reported on was 13.6 (2008, 14.1; 2003-9.8; 1998-10.9; 1993-24.3), whereas the average number of full-time application programmer and/or analysts was 4.2 (2008-2.7; 2003-4.1; 1998-4.6; 1993-9.5). The average number of full time hired IT consultants was 3.1 (2008 - 2.8; 2003-0.7; 1998-2.7). The dip in 2003 on the number of consultants reflects the limited activity at the time in the Norwegian consultant-market (and general), where all the major consultant-companies had to lay off hundreds of employees. An area first asked about in the 2008-investigation was the amount of outsourcing. Whereas only a few of the respondents reported to have outsourced all the IT-activities, 91% (2008-86.2%) of the organizations had outsourced parts of their IT-activity. As we see, the responding companies are of similar size as those we included in the previous investigation (2008). The average experience in the local IS-department was 8.7 years (2008-5.6; 2003-5.4; 1998-6.3; 1993-6.4) years.

The mean number of main systems in the organizations was 11.6 (2008-7.9; 2003-4.5; 1998-9.6; 1993-10.3). The mean user population of these systems was 92720 (2008- 4382; 2003– 314; 1998-498; 1993-541). It is in particular the number of external users that has increased a lot (also relative to the number of employees in the organizations); the average number of internal users of the systems was 852. The age distribution of the systems in our studies and the Swanson/Beath study is provided in Table 1. The average age of the systems was 5.8 years (2008 5.0; 2003-3.9; 1998-6.4; 1993-4.6; Swanson/Beath-6.6).

*Table 1: Age distribution of systems (percentage)*

| Age | 2013 | 2008 | 2003 | 1998 | 1993 | SB |
|---|---|---|---|---|---|---|
| 0-1 | 10 | 11 | 20 | 7 | 13 | 7 |
| 1-3 | 21 | 26 | 37 | 19 | 38 | 17 |
| 3-6 | 28 | 32 | 27 | 33 | 22 | 24 |
| 6-10 | 29 | 24 | 8 | 23 | 18 | 26 |
| > 10 | 12 | 7 | 8 | 18 | 9 | 26 |

In 1993, 58% of the systems were developed by the IS-organization, and only one percent was developed *in* the user organization. In 1998, however, 27% of the systems were developed by the IS-organization and 27% as custom systems *in* the user organization. In 2003 23 % of the systems are developed in the IS-organization, whereas in 2008 only 12% was developed in the IS organization. The number for 2013 (13%) is quite similar. The percentage of systems developed by outside firms is higher in the later studies (31% vs. 40% in 2008 vs. 35% in 2003, vs. 22% in 1998 vs. 12 % in 1993 vs. 15 % in Swanson/Beath). The percentage of systems developed based on packages with small or large adjustments is also comparatively high (42% vs. 40% in 2008 vs. 39% in 2003 vs. 24% in 1998 vs. 28% in 1993 vs. 2% in Swanson/Beath). The new category we introduced in 1998, component-based development (renamed "use of external web services" in 2008) is increasing with 11% in 2013 (6% in 2008, 1.0 % in 2003, 0.4% in 1998) of the total systems. Figure 2 illustrates the above figures on a high level, differentiating between packaged and custom development of main IT-applications in the portfolio, where we see how these lines have crossed in the last investigation, with more systems being packaged than custom made.

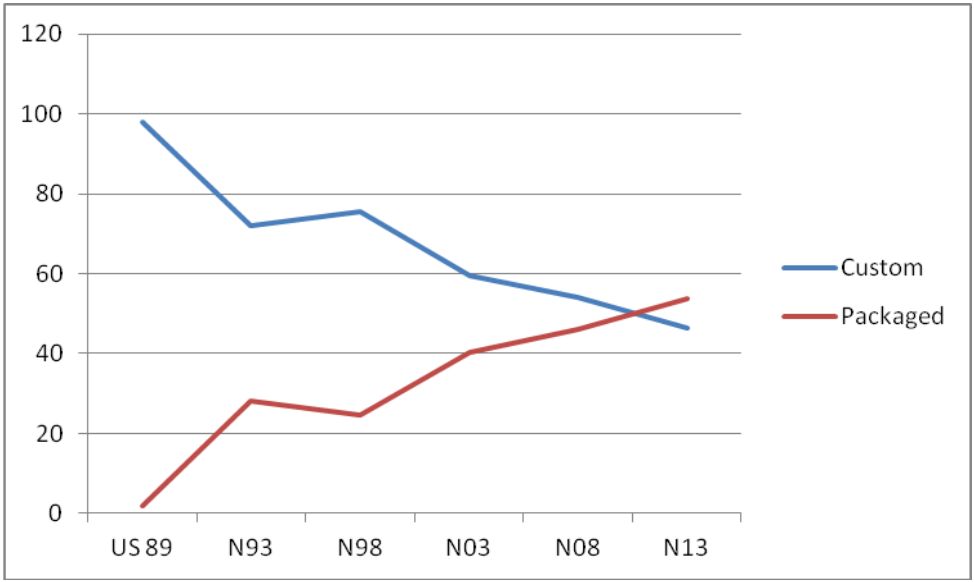

*Figure 2 Development of custom vs. packaged development*

The average number of different programming languages in use was 2.1 (median 2). This is similar to the investigations in 2008, 2003, 1998 and 1993. Table 2 provides an overview of the percentage of systems reported being developed using the different programming languages. As we see, from being dominant ten to fifteen years ago COBOL is almost not used in this century. Many other old languages are no longer used neither for main systems.

*Table 2: Use of different programming languages*

|  | 2013 | 2008 | 2003 | 1998 | 1993 | NP | LS |
|---|---|---|---|---|---|---|---|
| COBOL | 6.3 | 4.5 | 0.5 | 32.6 | 49 | 51 | 52 |
| 4GL | 12.5 | 12 | 13.5 | 16.9 | 24 | 8 |  |
| C | 4.2 | 2.4 | 12.5 | 15.4 | 4 | 3 |  |
| C++ | 12.5 | 17.5 | 23.1 | 15.1 |  |  |  |
| C# | 16.7 | 4.9 |  |  |  |  |  |
| Script-languages | 21.9 | 6.7 | NA | 12.9 | 4 | 10 | 22 |
| Java | 26 | 22.6 | 29.8 | 2 |  |  |  |
| Ass. |  | 0.4 |  | 0.9 | 3 |  | 12 |
| Fortran |  |  |  | 0.6 | 4 | 7 | 2.4 |
| PASCAL |  |  |  | 0.3 | 2 |  |  |
| PL/1 |  |  |  | 0.3 | 2 |  | 3.2 |
| Other | 12.5 | 28.9 | 20.2 | 2.6 | 6 | 21 | 7.7 |

The languages that are used in most organizations and for most systems in the last investigations are Java, C++ and this year C# and different script-languages. Java was just starting to be in widespread use in 1998 and C++ was barely included in 1993.

96 new systems were currently being developed, and as many as 58 of these systems (60 %) were regarded as replacement systems. (2008-66%; 2003-60%; 1998-57%; 1993-48%; SB-49%). The portfolio of the responding organizations responding to this question contained 749 systems, meaning that 13% of the current portfolio was being replaced. (2008 - 13%; 2003-13%; 1998-9%; 1993-11%; SB- 10%). The average age of systems to be replaced was 9.1 years (2008 - 6.8; 2003-5.5 years; 1998-7.7 years; 1993-8.5 years).

Table 3 summarizes reasons for the replacements, which have changed slightly from earlier investigations, but with a very similar pattern as in the previous investigation. The most important reasons for replacement are partly a need for integration/standardization and burden to maintain. One area which is expected to influence the software development and maintenance landscape is Service Oriented Architecture (SOA) (Lewis & Schmidt 2008). Transfer to SOA was very important as a reason to create replacement systems for only two organizations. 30% had started to implement SOA (2008 - 20%). Another technology development trend with particular relevance for easily providing new services and simplified systems operations is the use of cloud technology. So far only 5 % of the main systems in the respondent's application portfolios are supported by cloud technology.

*Table 3: Reasons for replacement of systems*

| Factor | Investigation | Order | Mean[1] |
|---|---|---|---|
| **Integration of systems** | 2013 | 1 | 3.4 |
| | 2008 | 1 | 3.7 |
| | 2003 | 1 | 3.4 |
| | 1998 | 2 | 3.2 |
| | 1993 | 1 | 3.9 |
| **Burden to maintain** | 2013 | 2 | 3.4 |
| | 2008 | 2 | 3.7 |
| | 2003 | 4 | 2.9 |
| | 1998 | 3 | 3.1 |
| | 1993 | 2 | 3.7 |
| | Swanson /Beath | 1 | 3.8 |
| **Standardization** | 2013 | 3 | 3.2 |
| | 2008 | 5 | 3.0 |
| | 2003 | 2 | 3.3 |
| | 1998 | 1 | 3.4 |
| | 1993 | 4 | 3.0 |
| **Burden to operate** | 2013 | 4 | 3.0 |
| | 2008 | 3 | 3.3 |
| | 2003 | 6 | 2.6 |
| | 1998 | 5 | 2.3 |
| | 1993 | 6 | 2.6 |
| | Swanson /Beath | 3 | 2.8 |
| **Burden to use** | 2013 | 5 | 2.9 |
| | 2008 | 6 | 2.9 |
| | 2003 | 6 | 2.6 |
| | 1998 | 7 | 2.1 |
| | 1993 | 4 | 3.0 |
| | Swanson/Beath | 1 | 3.8 |
| **New technical architecture** | 2013 | 6 | 2.7 |
| | 2008 | 4 | 3.1 |
| | 2003 | 3 | 3.0 |
| | 1998 | 4 | 2.9 |
| | 1993 | 2 | 3.7 |
| **Package alternative** | 2013 | 7 | 2.5 |
| | 2008 | 7 | 2.6 |
| | 2003 | 5 | 2.8 |
| | 1998 | 6 | 2.1 |
| | 1993 | 7 | 2.4 |
| | Swanson/Beath | 4 | 1.9 |
| **Generator alternative** | 2013 | 8 | 1.8 |
| | 2008 | 8 | 1.9 |
| | 2003 | 8 | 1.9 |
| | 1998 | 8 | 1.6 |
| | 1993 | 8 | 1.8 |
| | Swanson/Beath | 5 | 1.3 |

Work on application systems was in the survey divided into the six categories presented in section 2. The same categories were also used in 1993, 1998 and 2003 and 2008. We also asked for the time used for user-support and for systems operations which took up the additional time for the work in the IS departments. For these figures we have numbers both from the main replication study and the IT i praksis- study, and we present here the aggregated num-bers from these studies below. Note that in the surveys, we do not ask for numbers of our specific figures on application portfolio evolution and upkeep, but calculate them from figures of the more well-known types of maintenance and development.

---

[1] The use of a mean value here is only to have the possibility to get a crude comparison with the investigation by Swanson/Beath, which did not report the distribution. This number has been calculated by giving the value 5 to high importance, 4 for substantial importance etc.

Table 4 shows the distribution of work in previous investigations, listing the percentage of maintenance work, the study reported, and the year of the study. Based on this we find that in most investigations, between 50% and 70% of the effort is done to enhance and adapt systems in operation (maintenance) when disregarding other work than development and maintenance, although with a slightly increasing trend. An exception from this was our study in 1998 that was found to be influenced particularly by the amount of Y2K-oriented maintenance (Holgeid et. al. 2000). The numbers reported by Jones (2006) were also higher than this, but these also include user support as part of maintenance contrary to what we have done in our surveys. Dekleva (1992) also include user support as part of maintenance.

*Table 4: Result on percentage of maintenance from previous investigations*

| %maintenance | Investigation | Year |
|---|---|---|
| 49 | Arfa, Mili & Sekhri (1990) | 1990 |
| 53 | Lientz & Swanson (1980) | 1980 |
| 56 | Jørgensen (1994) | 1994 |
| 58 | Yip (1995) | 1995 |
| 59 | Krogstie & Sølvberg (1994) | 1993 |
| 62 | Nosek & Palvia (1990) | 1990 |
| 66 | Davidsen et al (2010a) | 2008 |
| 66 | Krogstie et al (2006) | 2003 |
| 66 | Dekleva (1992) | 1990 |
| 73 | Holgeid et al (2000) | 1998 |
| 73 | Capers Jones (2006) | 2000 |
| 79 | Capers Jones (2006) | 2010 est. |

Table 5 summarizes the descriptive results on the distribution of work in the categories in our investigation, comparing to previous investigations.

*Table 5: Distribution of the work done by IS-departments*

| Category | 2013 | 2008 | 2003 | 1998 | 1993 | LS |
|---|---|---|---|---|---|---|
| Corrective | 10.2 | 8.2 | 8.8 | 12.7 | 10.4 | 10.6 |
| Adaptive | 9.7 | 6.3 | 7.3 | 8.2 | 4 | 11.5 |
| Enhancive | 12.9 | 11.3 | 12.9 | 15.2 | 20.4 | 20.5 |
| Non-functional perfective | 8.0 | 9.1 | 7.6 | 5.4 | 5.2 | 6.4 |
| **Total maintenance** | **40.7** | **34.9** | **36.6** | **41,5** | **40** | **48.8** |
| Replacement | 8.3 | 9.7 | 9.9 | 7.7 | 11,2 | NA |
| New development | 8.3 | 11.4 | 12.5 | 9.5 | 18,4 | NA |
| **Total development** | **16.6** | **21.1** | **22.4** | **17.1** | **29.6** | **43.3** |
| Technical operation | 23.4 | 23.8 | 23.6 | 23 | NA | NA |
| User support | 19.2 | 20.1 | 17.2 | 18.6 | NA | NA |
| **Other** | **42.6** | **43.9** | **40.8** | **41.6** | **30.4** | **7,9** |

40.7% of the total work among the responding organizations is maintenance activities, and 16.6% is development activities. When disregarding other work than development and maintenance of application systems, the percentages are as follows: maintenance activities: 73%, development activities: 27 %. This is a bit more skewed towards maintenance than in the previous investigations, back to the level reported in 1998. 65% of development and maintenance work was application portfolio upkeep, and 35% was application portfolio evolution. This is almost the same as in 2008, 2003 and 1998, which in turn was significantly different from the situation in 1993 where application portfolio upkeep- and application portfolio evolution respectively amounted to 44% and 56% of the work (see table 4).

Table 6 summarizes the results on the breakdown of maintenance activities from our investigations where we look upon the complete portfolio of the responding organizations. Most interesting for comparison with other surveys is looking at corrective, adaptive, and perfective maintenance, which appears to be much more stable than the numbers reported from others. We do note though that the enhancive maintenance part of perfective maintenance appears to have stabilized on a lower level than we found 20 years ago.

*Table 6. Distribution of maintenance activities*

| Category | 2013 | 2008 | 2003 | 1998 | 1993 | LS |
|---|---|---|---|---|---|---|
|  | Mean | Mean | Mean | Mean | Mean | Mean |
| 1 Corrective maintenance | 25 | 24 | 24 | 31 | 26 | 17 |
| 2 Adaptive maintenance | 24 | 19 | 20 | 20 | 10 | 18 |
| 3 Enhancive maintenance | 32 | 30 | 35 | 37 | 51 |  |
| 4 Non-functional perfective maintenance | 19 | 27 | 21 | 13 | 13 |  |
| Perfective maintenance (3+4) | 51 | 57 | 56 | 50 | 64 | 60 |

Further comparisons of descriptive results between different studies are presented in Table 7. The first column lists the category, whereas the other columns list the numbers from our investigation, the one in 2003, the one in 1998, the one in 1993, the Nosek/Palvia (NP) investigation and the Lientz/Swanson (LS) investigation. The first set of number compare the numbers for development, maintenance and other work. The amount of *other work* reported in our investigations is much larger than in the American investigations. Therefore, in the second set of figures, we compare the data without considering other work. For the categories application portfolio evolution and application portfolio upkeep, we only have numbers from our own investigations.

*Table 7: Comparisons of maintenance figures with previous investigations*

| Category | 2013 | 2008 | 2003 | 1998 | 1993 | NP | LS |
|---|---|---|---|---|---|---|---|
| Development | 17 | 21 | 21 | 17 | 30 | 35 | 43 |
| Maintenance | 41 | 35 | 35 | 41 | 40 | 58 | 49 |
| Other work | 43 | 44 | 44 | 42 | 30 | 7 | 8 |
| *Disregarding other work* | | | | | | | |
| Development | 27 | 34 | 34 | 27 | 41 | 38 | 47 |
| Maintenance | 73 | 66 | 66 | 73 | 59 | 62 | 53 |
| *Functional effort* | | | | | | | |
| Application portfolio evolution | 35 | 36 | 39 | 38 | 56 | NA | NA |
| Application portfolio upkeep | 65 | 64 | 61 | 62 | 44 | NA | NA |

# 5 HYPOTHESES ON TRENDS IN WORK DISTRIBUTION

The following main hypotheses were formulated before the investigation to investigate the development of the different measures for distribution of work.

- H1: There is no difference between the percentage of time used for maintenance reported in our survey and what are reported in previous surveys. Rationale: When comparing the percentage of time used for maintenance activities in organizations earlier, we have found this to be stable on close to 40 percent of the overall time in our last three investigations (1998, 2003, and 2008), slowly declining. We would not expect this to be different in this survey.

- H2: There is no difference between the percentage of time used on development reported in our survey and what are reported in previous surveys. Rationale: When comparing the percentage of time used for development activities in organizations earlier, we have found this to be stable on close to 20 percent of the overall time in investigations both in the seventies, eighties, and nineties in both USA and Norway. We would not expect this to be different in this survey.

- H3: There is no difference between the breakdown of maintenance work (in corrective, adaptive, and perfective maintenance) in our survey and what are reported in previous surveys. Rationale: A number of investigations (also from the later years) reporting on the distribution of time among maintenance tasks as summarized in (Gupta et al, 2006) reports very different numbers for these distributions. On the other hand the scope of these investigations varies greatly. Whereas some look on single systems of numerous organizations and the whole portfolio of several organizations, other look only at one or a few applications in one organization. Since this distribution naturally will differ according to where the individual system is in the lifecycle (development, evolution, servicing, phase-out, closed (Rajlich & Bennett 2000)), this difference should be expected when only looking on individual systems. When averaging across a large number of application portfolios, each consisting of a number of systems of different maturity on the other hand, we would expect a more stable distribution. We will investigate this relative to the percentage of corrective, adaptive, and perfective maintenance reported.

- H4: There is no difference between the distribution of work among maintenance and development in our survey and what is reported in previous surveys when disregarding other work than development and maintenance. Rationale: Since the amount of other work than development and maintenance is taking up more time now than earlier, we found it beneficial also in the surveys in 1993, 1998, 2003 and 2008 to look at only the proportion between development and maintenance time. When only taking development and maintenance into account, the amount of time used for maintenance has been shown to be stable on 65-70% (i.e., 30-35% for development) since the late 90ties, raising from around 60 % in earlier investigations.

- H5: There is no difference between the distribution of application portfolio upkeep and application portfolio evolution in our survey and what are reported in previous surveys. Rationale: Since these numbers had been investigated only in our own investigations, we were eager to find if also they had the same stable distribution as the maintenance figure. They were significantly higher in 1998 than in 1993, but on the same level in 2008, 2003 and 1998, a pattern we would expect to continue.

Test of significance of differences of results between earlier investigations have been presented in (Davidsen & Krogstie 2010a, Holgeid et al 2000, Krogstie et al 2006). Similar tests in the 2008 investigations showed that there was no significant development in these numbers. Although also in this round the differences on many variables are relatively small, the larger number of responses makes it more possible to find if the differences actually are significant. Before looking for significant relationships to follow up the stated hypothesis related to trends, the variables used in the comparisons were tested for normality as illustrated in Table 8 and 10. We provide here data to test the distribution of the relevant variables from the 2013 and 2008 investigation. As indicated by the significant numbers (in boldface), data for a number of variables cannot be investigated as if they where normally distributed, since we in these cases must reject the null-hypothesis that the numbers are normally distributed, since either the Shapiro-Wilks (S-W Sign) and/or the Kolmogorov-Smirnov (Lilliefors-Sign.) significance levels are less than 0.05. On some variables (e.g. application portfolio upkeep) we could use the assumption of normal distribution in the tests below, using t-tests, whereas for the others we use a non-parametric test (Mann-Whitney).

*Table 8: Test for normality of variables*

| **Figure** | **Kolmogorov-Smirnov** | **Sign (p)** | **Shapiro-Wilks** | **S-W Sign (p)** |
|---|---|---|---|---|
| Corrective maintenance 2013 | .216 | **.000** | .904 | **.000** |
| Corrective maintenance 2008 | .208 | **.000** | .752 | **.000** |
| Adaptive maintenance 2013 | .176 | **.000** | .882 | **.000** |
| Adaptive maintenance 2008 | .231 | **.000** | .881 | **.000** |
| Perfective maintenance 2013 | .136 | **.005** | .880 | **.000** |
| Perfective maintenance 2008 | .080 | .200 | .973 | .204 |
| Maintenance 2013 | .092 | .200 | .979 | .340 |
| Maintenance 2008 | .073 | .200 | .986 | .705 |
| Development 2013 | .158 | **.000** | .864 | **.000** |
| Development 2008 | .106 | .086 | .946 | **.009** |
| Maintenance relative to development 2013 | .064 | **.000** | .871 | **.000** |
| Maintenance relative to development 2008 | .117 | **.038** | .941 | **.006** |
| Application portfolio upkeep 2013 | .093 | .200 | .966 | .051 |
| Application portfolio upkeep 2008 | .096 | .200 | .961 | .051 |

We tested H1-H5 by comparing with our previous survey as summarised in Table 9.

*Table 9: Test of hypothesis*

| | **Year** | **N** | **Mean** | **SD** | **Δ** | **P** |
|---|---|---|---|---|---|---|
| Maintenance, percentage of all work (vs. H1) | 2008 | 63 | 34.9 | 17.6 | 7.2 | **.008** |
| | 2013 | 66 | 42.1 | 19.7 | | |

| Figure | Year | N | Mean | SD | Δ | p |
|---|---|---|---|---|---|---|
| Development, percentage of all work (vs. H2) | 2008 | 63 | 21.1 | 16.4 | 8.3 | **.005** |
| | 2013 | 66 | 13.8 | 14.2 | | |
| Corrective maintenance, percentage of all work (vs. H3a) | 2008 | 63 | 8.2 | 8.0 | 1.2 | 428 |
| | 2013 | 66 | 9.4 | 6.7 | | |
| Adaptive maintenance, percentage of all work (vs. H3b) | 2008 | 63 | 6.3 | 5.5 | 3.7 | **.035** |
| | 2013 | 66 | 9.0 | 5.9 | | |
| Perfective maintenance, percentage of all work (vs. H3c) | 2008 | 63 | 20.1 | 13.7 | 3.6 | .453 |
| | 2013 | 66 | 23.7 | 16.9 | | |
| Maintenance, disregarding other work (vs. H4) | 2008 | 61 | 65.7 | 21.5 | 12 | **.001** |
| | 2013 | 64 | 77.7 | 20.0 | | |
| Application portfolio upkeep (vs. H5) | 2008 | 61 | 63.9 | 18,6 | 4.6 | .164 |
| | 2013 | 64 | 68.5 | 17.8 | | |

We list the number of cases, the mean and the standard deviation for all relevant figures to test the seven hypotheses (for H3, there are three test, for the difference in corrective, adaptive, and perfective maintenance respectively), having the numbers from 2008 on the top and those from 2013 on the bottom. Δ is the absolute difference in the mean from the 2008 and the 2013 study, and p is the probability for erroneously rejecting the equality of means.

We repeated the analysis taking also into account the responses from the IT i praksis - study. Below we include the normality test taking into account all data from 2013.

*Table 10: Test for normality of variables including IT i praksis*

| Figure | Kolmogorov-Smirnov | Sign (p) | Shapiro-Wilks | S-W Sign (p) |
|---|---|---|---|---|
| Corrective maintenance 2013 | .238 | **.000** | .770 | **.000** |
| Adaptive maintenance 2013 | .124 | **.000** | .922 | **.000** |
| Perfective maintenance 2013 | .121 | **.000** | .901 | **.000** |
| Maintenance 2013 | .086 | **.000** | .979 | **.001** |
| Development 2013 | .117 | **.000** | .271 | **.000** |
| Maintenance relative to development 2013 | .097 | **.000** | .948 | **.000** |
| Application portfolio upkeep 2013 | .051 | .089 | .990 | .060 |

We tested again H1-H5 by comparing with our previous survey as summarised in Table 6.

*Table 11: Test of hypothesis*

|  | Year | N | Mean | SD | Δ | P |
|---|---|---|---|---|---|---|
| Maintenance, percentage of all work (vs. H1) | 2008 | 63 | 34.9 | 17.6 | 6.9 | .053 |
|  | 2013 | 271 | 40.8 | 16.4 |  |  |
| Development, percentage of all work (vs. H2) | 2008 | 63 | 21.1 | 16.4 | 4.5 | **.035** |
|  | 2013 | 271 | 16.6 | 14.2 |  |  |
| Corrective maintenance, percentage of all work (vs. H3a) | 2008 | 63 | 8.2 | 8.0 | 2 | .175 |
|  | 2013 | 275 | 10.2 | 8.0 |  |  |
| Adaptive maintenance, percentage of all work (vs. H3b) | 2008 | 63 | 6.3 | 5.5 | 3.4 | **.000** |
|  | 2013 | 275 | 9.7 | 6.4 |  |  |
| Perfective maintenance, percentage of all work (vs. H3c) | 2008 | 63 | 20.1 | 13.7 | 0.8 | .884 |
|  | 2013 | 275 | 20.9 | 12.7 |  |  |
| Maintenance, disregarding other work (vs. H4) | 2008 | 61 | 65.7 | 21.5 | 7.5 | **.013** |
|  | 2013 | 271 | 73.2 | 19.1 |  |  |
| Application portfolio upkeep (vs. H5) | 2008 | 61 | 63.9 | 18,6 | 1.3 | .613 |
|  | 2013 | 271 | 65.2 | 16.4 |  |  |

Table 11 is built up in the same manner as table 9. Comparing table 9 and 11, we find the same differences being significant, with the exception of 'maintenance, percentage of all work'. We briefly summarise the results in the start of the conclusion.

## 5 CONCLUSION AND FURTHER WORK

Revisiting our hypotheses, we conclude the following:

- H1: There is no difference between the percentage of time used for maintenance reported in our survey and what are reported in previous surveys: Not rejected. From table 11, we observe a large increase in this number (from 34.9 to 41.3 percent), but this is not significant. Only looking at our data from the replication study, we find this to be significant, with 42.1 percent. We note from table 5 that we in both cases seem to have returned to the level reported in the nineties on this figure.

- H2: There is no difference between the percentage of time used on development reported in our survey and what are reported in previous surveys: Rejected. The amount of development work has gone down significantly (on the 0.05 - level) between 2008 and 2013, from 21.1 to 16.8% of the total time. In the replication study, it was reported as little as 13.8 percent, also a significant difference. The level of development in 2013 is as low as it was in 1998, which was largely influenced by the Y2K-preperations. It is in particular the time for development of new systems that is lower. This can be related to that more of the development work than the other types of work is being outsourced to external organizations, a hypothesis we will investigate in further work.

- H3: There is no difference between the breakdown of maintenance work (in corrective, adaptive, and perfective maintenance) in our survey and what are reported in previous surveys. Not rejected for corrected and perfective maintenance, but rejected for adaptive maintenance. A large number of changes in the underlying IT-development infrastructure such as the introduction of cloud computing and increasing popularity of mobile applications might be behind this development.

- H4: There is no difference between the distribution of work among maintenance and development in our survey and what is reported in previous surveys when disregarding other work than development and maintenance. Rejected. The percentage of development vs. maintenance is back at the level reported before Y2K (1998), having a significant increase in the amount of work done for maintenance.

- H5: There is no difference between the distribution of application portfolio upkeep and application portfolio evolution in our survey and what are reported in previous surveys. Not rejected. Even if more work is done on maintenance, more of this work is done on enhancive maintenance, offloading the decrease in the amount of work spent on developing new systems. Thus even if the report on maintenance figures only could be a source of alarm, the functional maintenance measure which we have argued gives a better indication on to what extent resources are used well in an organization, is kept at a stable, albeit arguably high level.

To explain differences better, we will look on other aspects from the replication study. There are for instance a number of differences in the underlying technology, which is as expected. This is very clearly witnessed in the distribution of programming languages used, where procedurally languages like COBOL to a large extend have been suppressed by object-oriented languages like Java, C++ and C# and by scripting languages. Note still that this has happened later than one might would have originally expected, applications exist for a number of years in organizations before they are being replaced. New architectural trends such as SOA and cloud computing are starting to make some impact on the use of resources, although less than expected given that it is more than ten years ago that web services and SOA was heralded as the way to develop systems for the future. Another marked trend is that less and less of IT is done internally in organizations (this applies to development, maintenance, operations and use). On the other hand, even if most organizations outsource part of the IT-activities, most still do the majority of the activities in-house. Overall percentage of time used for application portfolio evolution is remarkably stable. The same can be said about the rate of replacement, although slightly increasing, more than 60% of 'new' systems to be developed are actually replacement systems, constituting around 13% of the current application portfolio. Since more complex infrastructures are supporting the information systems, and they have increasing number of users, more of the resources are used for other tasks than development and maintenance than when we started these investigations, but also this seems to have stabilized.

As we in the current investigation have data from many more organizations than in the previous studies, it will hopefully be easier to find significant results relative to what characterizes organizations with good practice, having a high degree of application portfolio evolution. Among other things we investigate the differences between private and public sector organization. This follows up the previous exploratory investigation on this difference (Krogstie 2012). Here, small, but not statistically significant differences were found. Another area is to investigate to what extent the integration of IT and business strategy influences the ability to prioritize resources to developing new functionality either as new systems or as enhancive maintenance.

A survey investigation of this form has known limitations as discussed in section 3. To come up with more concrete empirical data on to what extent the application systems support in an organization is efficient, demands another type of investigation, surveying the whole portfolio of the individual organization, and getting more detailed data on the amount of the work that is looked upon as giving the end-user improved support, and how efficient this improved support was provided. This should include the views of the users of the application systems portfolio in addition to those of the IS-managers and developers. Results from such detailed case studies on the other hand are hard to generalize.

We have just started on the analysis of the data, and we are currently performing more detailed statistical analysis. Contrary to our main study, 'IT i praksis' is run yearly, and getting high level work distribution data regularly is interesting for further investigations. A long-term plan is to do a similar replication investigation in 2018, following up our five-year cycle, but before that have done some additional case studies to more precisely pinpoint relevant issues including methodological, managerial and technological trends to investigate.

## ACKNOWLEDGEMENT

We would like to thank all the participants of the survey-investigation for their effort in filling in the forms. We would also like to thank everyone helping us in the piloting and refinement of the questionnaire.

# APPENDIX A – Contents of the survey form

Below is listed the main questions from the survey form. This is not an exact copy of the form used. For reasons of brevity, we have only included the questions relevant to the results presented in the paper. We have below changed the layout and removed most of the room for giving additional information and qualification of the answers provided in the SurveyMonkey forms. We have neither included the additional material explaining the format and vocabulary used in the form. The survey form content has also been translated into English from Norwegian.

---

**4. Current position:** ___IT-Manager
___Project manager
___System developer, designer etc.

**6. Years of computer experience:** ____

**7. Type of organisation (Telecom, banking, etc.):** _____

**10. Number of employees in your organisation:** ____

**11 What is the annual budget of the IS-organisation in 2008 including hardware, software and personnel (in mill. NOK)?**
                                  2013
a. more than 50         ___
b. between 40 and 50    ___
c. between 30 and 40    ___
d. between 20 and 30    ___
e. between 10 and 20    ___
f. between 1 and 10     ___
g. less than 1          ___

**12. How much of the following activity is outsourced:**

a. ___ The total IT-activity (%)
b. ___ Development of new applications (%)
c. ___ Maintenance of existing application (%)
d. ___ Operations (%)
e. ___ User support (%)
f  ___ Other specify: _______________________

**14. Distribute your IS department's work into the following categories:**

%
a. ___Correcting errors in systems in operation
b. ___Adapt the system to changed technical architecture
c. ___Develop new functionality in existing systems
d. ___Improve non-functional properties (e.g. performance)
e. ___Develop new systems which provide similar functionality as existing systems
f. ___Develop new systems to cover new functional areas
g. ___Operation

h. ___User support
i. ___Other, specify: ____________________
Total: **100%**

**16 Your answer above is:**
a. ___Reasonable accurate, based on good data
b. ___A rough estimate, based on minimal data
c. ___A best guess, not based on any data

**17. Specify the number of full-time positions in the IS department?** _______ positions

**18. How many of these positions are dedicated to system developers?** _______ positions

**19. Specify the distribution of system developers with respect to their experience in the IS department?**
Years           Persons
0-1             _______
1-3             _______
3-6             _______
6-10            _______
> 10            _______

**20. What is the annual average number of hire IT consultants (converted to full-time personnel?**
_______ persons

**21. Specify the number of current main systems in your organisation** _______ systems

**24. What is the total number of end-users?**
Internal_______   External_______

**26. Specify age of the main systems (years since first installation)?**
Years           Systems
0-1             ______
1-3             ______
3-6             ______
6-10            ______
> 10            ______

**27. What is the distribution of development backgrounds of the current installed application system portfolio?**

| | |
|---|---|
| Developed by the IS department | ______ systems |
| Developed by the user organisation | ______ systems |
| Developed by outside firm | ______ systems |
| Package with large internal adjustments | ______ systems |
| Package with small internal adjustments | ______ systems |
| Solutions based on external web-services | ______ systems |

**28. Which programming languages are in use? Please specify number of systems developed in each programming language.**

| Language | Number of systems | |
|---|---|---|
| COBOL | _______ | |
| C | _______ | |
| C++ | _______ | |
| C# | _______ | |
| Java | _______ | |
| Scripts (PHP/Pearl etc) | _______ | |
| 4GL | _______ | Specify |
| Other | _______ | Specify |

**30. Do what extend is SOA used for the current man systems?**

**32. How many of the main systems are based on cloud technology?**

**35.**     **What is the number of systems currently being developed?**
        _______ systems

**36.**     **Of the total number of new systems currently under development, how many of these are replacement systems (for systems currently in the application system portfolio)?**
        _______ systems

**37.**     **What is the age distribution of the systems to be replaced?**
        Years

| | | |
|---|---|---|
| 0-1 | _______ | systems |
| 1-3 | _______ | systems |
| 3-6 | _______ | systems |
| 6-10 | _______ | systems |
| >10 | _______ | systems |

**38.**     **In the case of systems to be replaced in the current installed application system portfolio, what are the important reasons for replacement?**
        **(For each statement below, pick a score from 1 to 5, 5 being most important)**

| | | |
|---|---|---|
| a. | Excessive burden to maintain the system | ______ |
| b. | Excessive burden to operate the system | ______ |
| c. | Excessive burden to use the system | ______ |
| d. | Existence of application package alternative | ______ |
| e. | Existence of application generator alternative | ______ |
| f. | Changes in technical architecture (e.g.SOA) | ______ |
| g. | Standardisation with rest of organisation | ______ |
| h. | Integration with new or existing systems | ______ |
| i. | Other, specify: _____________________ | ______ |

| | |
|---|---|
| 5: | Extreme importance |
| 4: | Substantial importance |
| 3: | Moderate importance |
| 2: | Slight importance |
| 1: | No importance |